\documentclass [11pt]{article}

\usepackage{amsmath,amssymb,epsfig,cite,color,verbatim}
\usepackage{graphics,float}
\usepackage{overpic} 
\usepackage{slashed,bbold}
\usepackage{soul}

\numberwithin{equation}{section}

\title{Axial Anomaly in the $SU(N)$ Gauge Matrix Model}
\date{\today}
\author{Nirmalendu Acharyya$^1$\footnote{nirmalendu@iitbbs.ac.in}, Mahul Pandey$^{2}$\footnote{mpandey@stp.dias.ie} and Sachindeo~Vaidya$^3\footnote{vaidya@iisc.ac.in}$ \\
${}^1${\small School of Basic Sciences, Indian Institute of Technology Bhubaneswar, Jatni, Khurda, Odisha 752050, India}\\
${}^2${\small School of Theoretical Physics, Dublin Institute for Advanced Studies, Dublin 4, D04 C932, Ireland}\\
${}^3${\small Centre for High Energy Physics,  Indian Institute of Science, Bengaluru, 560012, India}\\
}
\begin{document}

\maketitle
\vspace*{-1cm} 
\begin{abstract}
The $SU(N)$ Yang-Mills matrix model admits self-dual and anti-self-dual instantons. When coupled to $N_f$ flavors of massless quarks, the Euclidean Dirac equation in an instanton background has $n_+$ positive and $n_-$ negative chirality zero modes. We show that the index $(n_+ - n_-)$ is equal to a suitably defined instanton charge. Further, we show that the path integral measure is not invariant under a chiral rotation, and relate the
 non-invariance of the measure to the index of the Dirac operator. Axial symmetry is broken anomalously, with the residual symmetry being a finite group. For $N_f$ fundamental fermions, this residual symmetry is $\mathbb{Z}_{2N_f}$, whereas for adjoint quarks it is $\mathbb{Z}_{4N_f}$.
\end{abstract}

\section{Introduction}

An anomaly occurs when a symmetry of a classical theory fails to survive in the corresponding quantum theory. The earliest example is the axial anomaly \cite{Adler:1969gk, Bell:1969ts}. In massless QCD the classical action has symmetries corresponding to the vector and axial rotations on a Dirac spinor $\Psi$, defined as
\begin{eqnarray}
U(1)_V: \Psi\rightarrow e^{i\alpha}\Psi,\quad\quad U(1)_A: \Psi \rightarrow e^{i\alpha\gamma^5}\Psi
\end{eqnarray}
respectively. In the quantum theory, the former survives while the latter is anomalously broken. The axial anomaly leads to the nonconservation of the axial current $j^\mu_A=\bar{\Psi}\gamma^\mu\gamma^5\Psi$
\begin{equation}
\partial_\mu j_A^\mu \sim \epsilon^{\mu\nu\rho\sigma}F_{\mu\nu}F_{\rho\sigma}.
\label{ch_anomaly}
\end{equation}
where $F_{\mu\nu}$ is the gauge curvature.

In this article we will demonstrate a surprising result: the quantum $SU(N)$ gauge matrix model coupled 
to $N_f$ massless quarks also exhibits an anomalous breaking of axial symmetry. We show this by solving 
for the zero modes of the Euclidean Dirac operator in the background of instanton gauge configurations, and 
evaluating its index. The index is related to a suitably defined instanton charge of the background 
configuration, giving us the matrix model version of the Atiyah-Singer index theorem. Finally, following Fujikawa \cite{Fujikawa:1979ay, Fujikawa:1980eg}, we relate the index to 
the non-invariance of the path-integral measure, which gives the anomaly. For the fermions in the 
fundamental representation of $SU(N)$ (i.e. quarks), there remains a residual $\mathbb{Z}_{2N_f}$ 
symmetry, while for adjoint fermions the residual symmetry is $\mathbb{Z}_{4N_f}$. We emphasise that our result holds for any finite $N \geq 2$.

The surprise here is that the quantum $SU(N)$ gauge matrix model, at least at first sight, is vastly different 
from $SU(N)$ gauge field theory. Even the structure of instanton configurations is very different. In the gauge field theory, distinct self-dual instantons are labelled by the instanton charge $\pi_3(SU(N)) = \mathbb{Z}$, whereas in the corresponding matrix model, they are labelled by a finite set of integers. 
We find that even though the number of distinct self-dual instantons is finite, it is enough to disturb the balance between left- and right-handed fermion zero modes, leading to the anomalous breaking of axial symmetry.

That anomalies can also occur in quantum mechanical situations has been known for some time now \cite{Elitzur:1985xj,Esteve:1986db,Jackiw1991} and can be traced to domain problems. The Hamiltonian $H$, being an unbounded operator, can only be defined on a dense domain ${\cal D}(H)$ of the Hilbert space. A symmetry is broken anomalously if the generator ${\cal J}$ of the symmetry transformation does not preserve ${\cal D}(H)$. More recently, anomalies in quantum mechanics have been shown to play an important role in the possible phase structure of $\theta$-vacua  in QCD \cite{Gaiotto:2017yup,Kikuchi:2017pcp}. In this work, we show the anomaly using Euclidean methods rather than domain issues.

The matrix model discussed here was first presented in \cite{Balachandran2014iya,Balachandran2014voa,Pandey:2016hat}, 
and has been shown to be an excellent candidate for an effective low-energy approximation of $SU(N)$ 
Yang-Mills theory on $S^3 \times \mathbb{R}$. In particular, a numerical investigation of its spectrum 
gave remarkably accurate predictions for the light glueball and hadron masses \cite{Acharyya:2016fcn,Pandey:2019dbp}. To serve as a correct low-energy approximation of Yang-Mills 
theory, however, this quantum-mechanical model must also exhibit the axial anomaly. Our present work 
provides this important conceptual support.

The article is organized as follows. In Section 2, we present a brief derivation of the 
gauge matrix model coupled to massless quarks. In Section 3, we discuss the instantons in the pure 
gauge matrix model, and show how a suitably defined instanton charge can be assigned to elementary 
self-dual instanton configurations. In Section 4, we solve for the zero modes of the Dirac operator in the 
instanton background, and compute the index of the Dirac operator using Callias' Index Theorem. In Section 5, 
we show that in the presence of the instanton, the fermion path integral measure is not invariant under 
axial transformations, and thus the axial symmetry is broken anomalously. In Section 6, we couple 
adjoint fermions to the matrix model and show that the axial symmetry is broken in this situation as well. In 
Section 7 we discuss the implications of our result.

\section{The Matrix Model}
We briefly recall the construction of the Yang-Mills matrix model, first discussed in \cite{Balachandran2014iya,Balachandran2014voa}. The spatial $\mathbb{R}^3$ is compactified to $S^3$, which is then isomorphically mapped to an $SU(2)$ subgroup of the gauge group $SU(N)$.  We consider the general left-invariant Maurer-Cartan one-form $\Omega$ on $SU(N)$,
\begin{equation}
\Omega =\text{Tr} (T_a g^{-1}dg)M_{ab}T_b,\quad a,b= 1,\cdots,N^2-1
\end{equation}
where $M$ is a $(N^2-1)$-dimensional real matrix, $g \in SU(N)$ and  $T_a$ are the generators of $SU(N)$ in the fundamental representation, satisfying $\text{Tr} \,T_a T_b=\frac{1}{2}\delta_{ab}$.

Under the mapping $S^3 \to SU(2) \subset SU(N)$, the orthonormal basis of vector fields on $S^3$ can 
be identified with $i X_i$, where $X_i$ is the left-invariant vector field on $SU(2)$. { The gauge 
fields $A_i \equiv M_{ia} T_a$ ($i=1,2,3$) are the pullback of the Maurer-Cartan form $\Omega$ under this map on $S^3$.}
Here $M_{ia}$ is a $3 \times (N^2-1)$-dimensional real matrix depending on time only. Under spatial rotations $R \in SO(3)$ and gauge transformations $h \in SU(N)$, it transforms as
\begin{eqnarray}
 M_{ia} \rightarrow M'_{ia} = R_{ij}M_{ja}, \quad\quad M_{ia} \rightarrow M'_{ia} = M_{ib} S(h)_{ab}, 
\label{gaugetr1}
\end{eqnarray}
where $S(h) \in \text{ad}\,SU(N)$ is the image of $h$ under the adjoint map $h T_a h^{-1} = S(h)_{ba}T_b$. 
Equivalently, $A_i$ transforms via conjugation under gauge transformations $A_i\rightarrow A'_i= hA_i h^{-1}.$
The matrix model degrees of freedom are elements of $\mathcal{M}_N$, the set of all 
$3 \times (N^2-1)$-dimensional real matrices $M_{ia}$, and gauge transformations act as $M \to M S(h)^T$, 
$h \in SU(N)$. The configuration space of the model is the base space of principle bundle $\text{Ad} \,SU(N)\rightarrow\mathcal{M}_N\rightarrow\mathcal{M}_N /\text{Ad}\, SU(N)$. This fibre bundle has been previously studied and found to be twisted, not admitting global sections \cite{Singer:1978dk,Narasimhan:1979kf}, and this fact lies at the heart of the Gribov problem in Yang-Mills theory. 

The curvature $F_{ij}$ is obtained by the pullback of the structure equation $d\Omega + \Omega \wedge \Omega$ to $S^3$:
\begin{eqnarray}
 F_{ij} 
&=&-\epsilon_{ijk}A_k-i[A_i,A_j].
\end{eqnarray}
For dynamics, we need a gauge-invariant Lagrangian. The obvious choice for the kinetic term 
$\frac{1}{2}\dot{M}_{ia}\dot{M}_{ia}=\text{Tr}(\dot{A}_i\dot{A}_i)$ is not gauge-invariant because the time-derivative of $A_i$ does not transform covariantly under time-dependent gauge transformations. Rather,
\begin{equation}
\dot{A'_i}= h(\dot{A}_i+[h^{-1}\dot{h},A_i])h^{-1}.
\end{equation}
To remedy this, we introduce a set of time-dependent real functions, conveniently named as $M_{0a}$, and the matrix $A_0 \equiv M_{0a}T_a$. We then define the gauge-covariant time derivative of $A_i$ as
\begin{equation}
D_t A_i =\dot{A}_i-i[A_0, A_i].
\end{equation}
Under a gauge transformation $h(t)\in SU(N)$, $D_t A_i$ transforms covariantly, i.e., $D_t A_i\rightarrow h(D_t A_i)h^{-1}$, provided the matrix $A_0$ transforms as
\begin{equation}
A_0\rightarrow A'_0=hA_0h^{-1}-\dot{h}h^{-1}.
\label{gaugetr2}
\end{equation}
$A_0$ is therefore the parallel-transporter needed to define the covariant derivative along the temporal 
direction. 
The electric field $E_i \equiv D_t A_i$ and the magnetic field $B_i \equiv \frac{1}{2}\epsilon_{ijk}F_{jk}$ are given by
\begin{eqnarray}
E_i 
= \dot{A}_i-i[A_0,A_i],\quad\quad 
B_i
=-A_i-\frac{i}{2}\epsilon_{ijk}[A_j,A_k].
\end{eqnarray}

We choose the dynamics of the gauge fields to be governed the Yang-Mills Lagrangian
\begin{equation}
L_{YM}=\frac{1}{g^2}\text{Tr}(E_iE_i-B_iB_i) = \frac{1}{g^2}\text{Tr}(D_tA_iD_tA_i)-V,
\label{LYM}
\end{equation}
where the potential { $V(A)\equiv \frac{1}{g^2}\text{Tr}\,B_iB_i$ consists of quadratic, cubic and quartic interaction terms: 
\begin{eqnarray}
V(A)
&=& \frac{1}{g^2}\text{Tr} \left( A_i A_i +i \epsilon_{ijk} [A_i, A_j] A_k -\frac{1}{2} [A_i, A_j][A_i,A_j]\right). 
\label{pot}
\end{eqnarray}}
We have set the radius $R$ of $S^3$ to 1, it can easily be put back when necessary. 

The model described above has been used to make variational estimates of the glueball spectrum \cite{Acharyya:2016fcn}, and has reproduced the predictions of lattice gauge theory with surprising accuracy.

Quarks are easy to introduce in the matrix model \cite{Pandey:2016hat,Pandey:2019dbp}. These are spinors $\Psi_{\alpha A}$, where $A$ and $\alpha$ denote the color and spin indices respectively. $\Psi_{\alpha A}$ are Grassmann-valued matrices that depend only on time, and transform in the fundamental representation of color $SU(N)$ and in the spin-$\frac{1}{2}$ representation of spatial rotations:
\begin{align}
&\Psi_{\alpha A } \rightarrow u(h)_{AB} \Psi_{\alpha B}, \quad h \in SU(N)\\
&\Psi_{\alpha A } \rightarrow D^{\frac{1}{2}}(R)_{\alpha\beta}\Psi_{\beta A}, \quad R \in SO(3)
\end{align}
respectively, where $u(h)$ is the fundamental representation of $h$ and $D^{\frac{1}{2}}(R)$ is the spin-$\frac{1}{2}$ representation of $R$.

The time derivative $\dot{\Psi}$ does not transform covariantly under time-dependent gauge transformations, but $D_t \Psi = (\partial_t -i A_0)\Psi$ does. Thus the Lagrangian with minimally coupled massless quarks is given by
\begin{equation}
L=L_{YM}+L_F,
\end{equation}
where the fermionic part is given by the gauge-covariant Dirac Lagrangian on $S^3$ \cite{Pandey:2016hat}:
\begin{eqnarray}
L_F = \bar{\Psi}\left( i \gamma^0 D_t +\gamma^iA_i -\frac{3}{2} \gamma^5 \gamma^0 \right) \Psi, \quad \bar{\Psi}\equiv\Psi^\dagger\gamma^0.
\label{LF}
\end{eqnarray}
In the chiral basis, the $\gamma$-matrices take the form
\begin{equation}
\gamma^0= \left( \begin{array}{cc}
0 & \mathbb{1} \\
\mathbb{1} & 0
\end{array}\right),\quad
\gamma^i= \left( \begin{array}{cc}
0 & \sigma_i \\
-\sigma_i & 0
\end{array}\right),\quad
\gamma^5= \left( \begin{array}{cc}
-\mathbb{1} & 0 \\
0 & \mathbb{1}
\end{array}\right),
\end{equation}
where $\mathbb{1}$ is the identity matrix in two dimensions and $\sigma^i$ are the Pauli matrices.

It is now straightforward to perform a Wick rotation $t\rightarrow -i\tau$ and $A_0\rightarrow i A_0$ in 
the Lagrangian $L$ \cite{gpy, LV}. Integrating $L$ over Euclidean time $\tau$ gives us the Euclidean action { $S_E  = S_E^{YM}+S_E^F$:
\begin{eqnarray}
S_E^{YM} &=& \frac{1}{g^2}\int d\tau \,\, \text{Tr} \left[ \left(\frac{ \partial A_i}{\partial  \tau}-[A_0,A_i]\right)^2 
+  A_i^2+ i \epsilon_{ijk} [A_i A_j] A_k
-  \frac{1}{2}[A_i, A_j]^2
\right], \label{seucym} \\
S_E^F &=& \int d\tau  \,\bar{\Psi} (i \slashed{D}) \Psi, 
\end{eqnarray}
where 
$D_\tau = \frac{\partial}{\partial\tau}- A_0$ and the Euclidean Dirac operator $\slashed{D}$ is given by 
\begin{eqnarray}
\slashed{D}= -i \left(\gamma^0 \partial_\tau  -\gamma_0A_0   -  \gamma^i A_{i}  +\frac{3}{2} \gamma^5 \gamma^0\right).
\label{diracop}
\end{eqnarray}

The Euclidean action $S_E$ is invariant under $U(1)_A$ axial rotation $\Psi \to e^{ i  \alpha \gamma^5} \Psi$. 
However, as we shall see, quantum effects break the axial $U(1)_A$ symmetry.

\section{Instantons in $SU(N)$ matrix model}

Classical vacuum configurations of the pure Yang-Mills matrix model are given by those $A_i$ for which $V(A)=0$. From (\ref{pot}), it is easy to see that the vacuum configurations satisfy
\begin{equation}
[A_i,A_j]=i\epsilon_{ijk}A_k.
\end{equation} 
This has solutions $A_i = 0$, or $A_i = L_i$, where $L_i$ is a generator of the Lie algebra of $SU(2)$. Thus classically, the matrix model has multiple degenerate vacua, which correspond to the matrices $A_i$ forming a general $N$-dimensional representation of the generators of $SU(2)$.

The existence of degenerate classical minima naturally opens the possibility of quantum mechanical tunnelling between them. The tunnelling behaviour is captured by  instantons, the finite action solutions of equations of motion of the action (\ref{seucym}). Defining the Euclidean chromoelectric field as
\begin{equation}
{\mathbb E}_i\equiv {\mathbb E}_i^aT_a = \frac{\partial A_i}{\partial \tau}-[A_0,A_i],
\end{equation}
we can rewrite the Euclidean action as
\begin{equation}
S_E^{YM}=\frac{1}{g^2}\int d\tau \text{ Tr}({\mathbb E}_i {\mathbb E}_i + B_i B_i) = \frac{1}{g^2}\int d\tau \text{ Tr}[({\mathbb E}_i \pm B_i)^2\mp 2 {\mathbb E}_i B_i]
\end{equation}
where $B_i = B_{ia}T_a = -A_i -\frac{i}{2}\epsilon_{ijk}[A_j,A_k] $. This action is extremized when
\begin{equation}
{\mathbb E}_i = \pm B_i.
\label{selfd}
\end{equation}
The solutions to (\ref{selfd}) with the plus (minus) sign give the (anti-)self-dual instantons. 

The quantity 
\begin{equation}
\int_{-\infty}^{\infty}d\tau \text{Tr} ({\mathbb E}_i B_i) = \int_{-\infty}^{\infty}d\tau \frac{d}{d\tau} \text{ Tr}\left( - \frac{1}{2} A_i A_i -\frac{i}{6}\epsilon_{ijk}A_i[A_j,A_k] \right)
\end{equation}
is the integral of a total derivative, is insensitive to localized $\tau$ fluctuations and hence an invariant.

We make use of our gauge freedom in { (\ref{gaugetr1}) }and (\ref{gaugetr2}) to transform (\ref{selfd}) to the temporal gauge, with $A_0=0$ \cite{Coleman:1985rnk}. 
Then the instanton equation (\ref{selfd}) becomes into
\begin{equation}
\frac{dA_i}{d\tau}=\pm \left(-A_i-\frac{i}{2}\epsilon_{ijk}[A_j,A_k]\right).
\label{instemporal}
\end{equation}

We can reverse this process and start with any solution of equation (\ref{instemporal}). 
A solution with $A_0\neq 0$ is obtained by considering all possible curves $h(\tau)$ in $SU(N)$ that takes finite values at the boundaries $\tau=\pm\infty$. For each such curve $h$, the instanton solutions are
\begin{align}
&A_i^h = hA_ih^{-1},\quad\quad A_0^h=-\frac{dh}{d\tau}h^{-1}.
\end{align}
The equations (\ref{selfd},\ref{instemporal}) and their solutions have been studied by \cite{Kronheimer:1990ay}, and also \cite{Bachas:2000dx} in a different context.

Substituting the ansatz 
\begin{equation}
A_i = \phi(\tau) L_i^{(1)} + \big(1-\phi(\tau)\big) L_i^{(2)}, \quad [L_i^{(1)}, L_j^{(2)}]=0 
\label{ansatz}
\end{equation}
into (\ref{instemporal}), where $L_i^{(\alpha)}$'s are generators of $SU(2)$ (in an arbitrary  $N$-dimensional representation), we obtain
\begin{eqnarray}
\frac{\partial \phi_s}{\partial \tau} =  -\phi_s (1- \phi_s), \quad\quad \frac{\partial \phi_a}{\partial \tau} =  \phi_a (1- \phi_a),
\end{eqnarray}
which have solutions
\begin{eqnarray}
\phi_s (\tau)= \frac{1}{1 + e^{(\tau -\tau_0)}},\quad\quad\phi_a(\tau) = \frac{1}{1 + e^{-(\tau -\tau_0)}}. 
\end{eqnarray}
The subscripts $s$ and $a$ denote self-dual and anti-self-dual solutions respectively. Their profiles are plotted in Fig. \ref{fig1}).
\begin{figure}[!hbtp]
\centering
\includegraphics[scale=1.1]{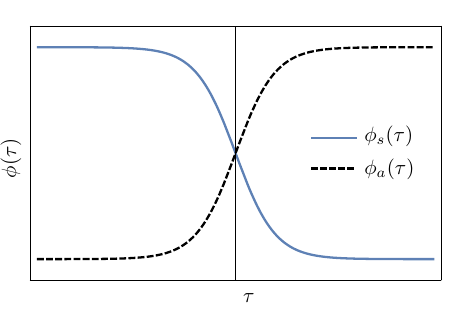}
\caption{$\phi_{s/a}(\tau)$ versus $\tau$ for $\tau_0=0$.}
\label{fig1}
\end{figure}

At $\tau = \pm \infty$, $\phi_{s/a}$ take finite values
\begin{eqnarray}
\phi_s (-\infty)= 1,\quad\quad\phi_s(\infty) = 0,\label{fsb}\\
\phi_a (-\infty)= 0,\quad\quad\phi_a(\infty) = 1.
\label{fab}
\end{eqnarray}

The self-dual instanton tunnels from the classical minimum at $A_i=L_i^{(1)}$ to $A_i = L_i^{(2)}$, and the anti-self-dual instanton from $A_i=L_i^{(2)}$ to $A_i=L_i^{(1)}$.

If we transform the solutions by the curve $h(\tau)$ to obtain $A_\mu^h$, we obtain instanton solutions that go between $L_i^{(1,2)}$ and $h_b L_i^{(2,1)} h_b^{-1}$, where $h_b$ is the value of $h$ at the boundary, $h_b=h(-\infty)$ for the self-dual and $h_b=h(\infty)$ for the anti-self-dual solution. This does not affect any of the subsequent arguments, since quantities of interest like the instanton number and charge defined below are gauge invariant. 

Up to a normalization factor $c$, the instanton number $\mathcal{T}$ is given by
\begin{equation}
\mathcal{T}=c \int_{-\infty}^\infty d\tau \text{ Tr } {\mathbb E}_i B_i= c \int_{-\infty}^{\infty}d\tau \frac{d}{d\tau} \text{ Tr}\left( - \frac{1}{2} A_i A_i -\frac{i}{6}\epsilon_{ijk}A_i[A_j,A_k] \right).
\end{equation} 
For the instanton (\ref{ansatz}),
\begin{align}
\mathcal{T} &= \pm \frac{c}{6}\Big(\text{Tr } L^{(2)}_iL_i^{(2)} -\text{Tr } L^{(1)}_iL_i^{(1)}\Big).
\label{314}
\end{align}

In (\ref{314}), if $L^{(\alpha)}_i$ is decomposed into a direct sum of $r_0^{(\alpha)}$ irreps, each irrep with $j_r^{(\alpha)} = \frac{N_r^{(\alpha)}-1}{2}$, then each block contributes $N_r^{(\alpha)} \frac{(N_r^{(\alpha)}-1)}{2} \frac{(N_r^{(\alpha)}+1)}{2}$. 
We can fix $c=4$ by requiring that $\mathcal{T}$ takes integer values for all cases. Thus
\begin{eqnarray}
\mathcal{T}&=&\pm \frac{1}{6} \left(\sum_{r=1}^{r_0^{(2)}} N_r^{(2)} (N_r^{(2)}-1) (N_r^{(2)}+1) - \sum_{r=1}^{r_0^{(1)}} N_r^{(1)} (N_r^{(1)}-1) (N_r^{(1)}+1) \right) \\
&=& \pm \frac{2}{3}\left(\sum_{r=1}^{r_0^{(2)}} j_r^{(2)} (j_r^{(2)}+1)(2j_r^{(2)}+1)- \sum_{r=1}^{r_0^{(1)}} j_r^{(1)} (j_r^{(1)}+1)(2j_r^{(1)}+1) \right).
\label{instantoncharge}
\end{eqnarray}

We could define $\mathcal{T}$ to be the instanton charge; however, the form (\ref{instantoncharge}) is not very enlightening as there is no simple relation between the $\{j_r^{(\alpha)}\}$ and $N$. Luckily, for the (anti-)self-dual 
instanton, we can construct a charge that is expressible only in terms of $r_0^{(\alpha)}$, the number of irreps at $\tau = \pm \infty$. 

We define ${\mathbb E'}_i$ as
\begin{equation}
{\mathbb E'}_i = \frac{d \phi_{s/a}}{d \tau}(e_i^{(2)} - e_i^{(1)}), \quad e_i^{(\alpha)} = \bigoplus_{r=1}^{r_0^{(\alpha)}}\frac{3}{(j_r^{(\alpha)}+1)(2j_r^{(\alpha)}+1)}L_i^{(\alpha),r} \label{Erescaled}
\end{equation}
where $L_i^{(\alpha),r}$ are the irreducible blocks with spin $j_r^{(\alpha)}$ in $L_i^{(\alpha)}$.

The new charge $4\int d\tau \, \text{ Tr } {\mathbb E'}_i B_i$ gives
\begin{eqnarray}
\mathcal{T}_{\text{new}}=4\int d\tau \, \text{ Tr } {\mathbb E'}_i B_i =  \pm(r_0^{(2)}-r_0^{(1)}). 
\label{topcharge}
\end{eqnarray}
The charge $\mathcal{T}_{\text{new}}$ is still an integral over a total $\tau$-derivative and hence a topological invariant. It depends only on $(r_0^{(2)}-r_0^{(1)})$, rather than the Casimirs or other labels of each individual irrep.
As we will demonstrate in the next section, $\mathcal{T}_{\text{new}}$ for a (anti-)self-dual instanton is {\it equal} to the index of the Dirac operator in that instanton background.

\section{Index of $\slashed{D}$ in Instanton Background}

The Dirac operator (\ref{diracop}) is Hermitian and obeys $\{\slashed{D}, \gamma^5\}=0$. 
So for every eigenfunction $\psi_n$ of $\slashed{D}$ with a non-zero eigenvalue $\lambda_n$,
there is an eigenfunction $\gamma^5 \psi_n$ with eigenvalue $-\lambda_n$.
This is not necessarily true for the eigenfunctions with zero eigenvalue (the zero modes). Since the Dirac operator 
is Hermitian, for non-zero eigenvalues, $\psi_n$ and $\gamma^5\psi_n$ are orthogonal. 

The zero modes can also be arranged to be eigenfunctions of $\gamma^5$: 
\begin{eqnarray}
\slashed{{D}} \chi_{k}^{ \pm} =0, \quad\quad \gamma^5 \chi_{k}^\pm = \pm \chi_{k}^\pm,\quad\quad k=1,2,\ldots n_\pm. 
\end{eqnarray}
$\chi_k^+$ and $\chi_k^-$ are the positive and negative-chirality zero modes. Let the number of zero modes with positive and negative chirality be $n_+$ and $n_-$ respectively. The index of the Euclidean Dirac operator $\slashed{D}$ is defined as the difference 
\begin{equation}
\text{ind}\,\slashed{D} = n_+ -  n_-.
\end{equation}

 In the Weyl basis, the Euclidean Dirac operator (\ref{diracop}) is 
 \begin{eqnarray}
\slashed{D} = 
 \left( \begin{array}{cc}
0 & \mathcal{L} \\
\mathcal{L}^\dagger &0
\end{array}\right),
\label{lmatrix} \quad\quad  
\mathcal{L} = -i\left(\frac{d}{d\tau} -A_0 + \sigma_i A_i + \frac{3}{2}\right).
\end{eqnarray}
The zero modes (in the temporal gauge) are fundamental Dirac spinors of the form $\Psi_{A\alpha}$. When $A_0^h = -\frac{dh}{d\tau}h^{-1}$, the corresponding zero mode is obtained via the gauge transformation $\Psi^h = u(h)\Psi$. In any case, the number of zero modes and hence the index do not change under the gauge transformation. So we can make our entire argument in temporal gauge.

Let us study the simple case when $A_i=\phi_{s/a}(\tau)L_i$, i.e. $L_i^{(2)}=0$ and $L_i^{(1)} = L_i$. Then the matrix { $\sigma_i A_i$} 
is a $2N$-dimensional Hermitian matrix whose $\tau$-dependence is via  $\phi_{s/a}(\tau)$ only. Its eigenvectors are $\tau$-independent,  so $\mathcal{L}$ can be diagonalized and $\slashed{D}$ brought to the form
\begin{eqnarray}
 \slashed{D}=  \left( \begin{array}{cc}
0 & \mathcal{L}_d \\
\mathcal{L}_d^\dagger &0
\end{array}\right), \quad \mathcal{L}_d = -i\left(\frac{d}{d\tau} + \Sigma\right),\quad \Sigma = \text{diag} ( \xi_1 (\tau), \xi_2 (\tau), \ldots \xi_{2N} (\tau)).
\label{lineq}
\end{eqnarray}
Here, $\xi_i (\tau)$'s are the eigenvalues of $\phi_{s/a} (\tau) \sigma_i \otimes L_i+\frac{3}{2}\mathbb{1}$.
Positive and negative chirality zero modes are of the form
\begin{equation}
\chi^+ = \left( \begin{array}{cc}
0 \\
\psi^+
\end{array}\right), \quad \chi^- = \left( \begin{array}{cc}
\psi^- \\
0
\end{array}\right)
\end{equation}
respectively, where the $\psi^{\pm}$ satisfy
\begin{equation}
\mathcal{L}_d\psi^+=0,\quad \mathcal{L}_d^\dagger\psi^-=0.
\end{equation}
The index can therefore be expressed as
\begin{equation}
\text{ind }\slashed{D} = \text{dim Ker}(\mathcal{L}_d)-\text{dim Ker}(\mathcal{L}^\dagger_d).
\end{equation}
We can find the $\psi^\pm$ easily from (\ref{lineq}): corresponding to the eigenvalue $\xi_n(\tau)$ of $\Sigma$, we obtain the zero mode
\begin{equation}
\psi_n^{\pm}(\tau) =  \text{ exp}\left(\mp \int^\tau d\tau' \xi_n(\tau')\right) C^\pm_n,\quad\quad n=1,2,\ldots 2N,
\label{exponential}
\end{equation}
where 
$C^\pm_n$ is a constant column vector with a non-zero entry only in the $n^{th}$ row.   

Let us count the \textit{normalizable} solutions for a given $L_i$.
Say $L_i$ consists of $r_0$ irreducible blocks (if $L_i$ is the $N$-dimensional irrep, $r_0=1$), each block of dimension $N_r\equiv (2j_r+1)$. The eigenvalues are then straightforward to calculate using addition of angular momenta: in each irreducible block,  $\phi_{s/a} (\tau) \sigma_i \otimes L_i$ has two distinct eigenvalues $ -\phi_{s,a}(\tau)\frac{N_r+1}{2}$ and $ \phi_{s,a}(\tau)\frac{N_r-1}{2}$  with multiplicities $N_r-1$ and $N_r+1 $ respectively. In total, $\Sigma$ has $(N_r-1)$ degenerate eigenvalues { $ \lambda_{1,s/a}  \equiv \frac{1}{2}[3-\phi_{s/a}(\tau) (N_r+1)]$}, and $(N_r+1)$ degenerate eigenvalues { $ \lambda_{2,s/a} \equiv\frac{1}{2}[3+\phi_{s/a}(\tau) (N_r-1)]$}. 

{ For either of these eigenvalues, the integral in (\ref{exponential}) evaluates to
\begin{eqnarray}
 \int^\tau d\tau' \xi_n(\tau') &=&\frac{3}{2}\tau-\frac{(N_r+ 1)}{2} \int^\tau d\tau' \phi_{s/a}(\tau'),\quad\quad \textrm{for }\xi_n = \lambda_{1,s/a}\\
 &=&\frac{3}{2}\tau+\frac{(N_r- 1)}{2} \int^\tau d\tau' \phi_{s/a}(\tau'),\quad\quad \textrm{for }\xi_n = \lambda_{2, s/a}
\end{eqnarray}
where (choosing  $\tau_0=0$ for convenience)
\begin{equation}
\int^\tau d\tau'\phi_{s}(\tau')=\tau-\text{ln}(1+e^\tau),\quad \int^\tau d\tau'\phi_{a}(\tau')=\text{ln}(1+e^\tau). 
\end{equation}}

Let us denote by $\psi_{s,n}^{\pm}$ the Dirac zero mode in the self-dual instanton background. For a given $N_r$,  there are $2N_r$   zero modes of the form
\begin{eqnarray}
\psi^\pm_{s,n}= \left\{ \begin{array}{lll}
e^{\mp\frac{1}{2}(2-N_r)\tau}(1+e^\tau)^{\mp\frac{N_r+1}{2}}  C^\pm_{s,n} , \quad \textrm{for } n=1,2, \ldots N_r-1  \\ \\
e^{\mp\frac{1}{2}(2+N_r)\tau}(1+e^\tau)^{\pm\frac{N_r-1}{2}} C^\pm_{s,n} , \quad \textrm{for } n=N_r, N_r+1 \ldots 2N_r. 
\end{array} \right. \label{sol2}
\end{eqnarray}

Similarly in the anti-self-dual instanton background 
\begin{eqnarray}
\psi^\pm_{a,n}= \left\{ \begin{array}{lll}
e^{\mp\frac{3\tau}{2}}(1+e^\tau)^{\pm\frac{N_r+1}{2}}  C^\pm_{a,n} , \quad \textrm{for } n=1,2, \ldots N_r-1  \\ \\
e^{\mp\frac{3\tau}{2}}(1+e^\tau)^{\mp\frac{N_r-1}{2}} C^\pm_{a,n} , \quad \textrm{for } n=N_r, N_r+1 \ldots 2N_r. 
\end{array} \right. \label{sol4}
\end{eqnarray}
To determine the index, we count the number of normalizable solutions for each chirality 
and compute $(n_+ -n_-)$. For example with $N_r =3$, the explicit plots of the solutions are shown in { Fig.\ref{fig2}a and Fig.\ref{fig2}b}.
\begin{figure}[!h]
\centering
\includegraphics[scale=1.1]{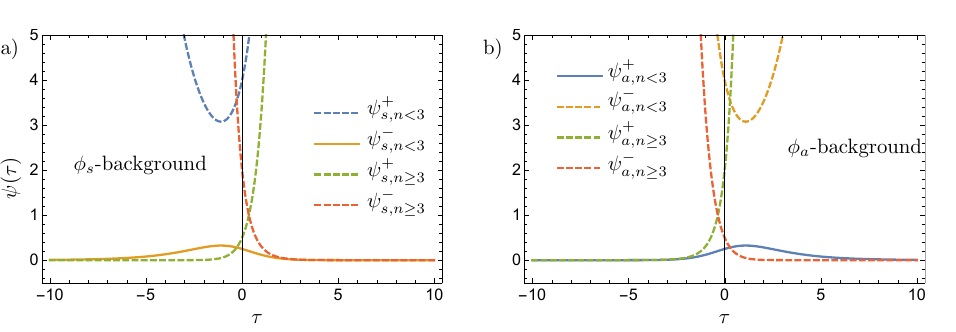}
\caption{Zero modes of  $\slashed{D}$ for  $N_r=3$ in a) a self-dual instanton, and b)  an anti-self-dual instanton background. The dashed lines are the non-normalizable solutions while the solid lines are the normalizable solutions. 
}
\label{fig2}
\end{figure}
In general there are $N_r-1$ normalizable solutions of with positive and negative chirality for self-dual and anti-self-dual instantons respectively.

We could, of course, analyze the time-dependence and normalizability of each of these solutions case by case; however, for a Dirac operator of the form (\ref{lmatrix}), there is a much neater way to determine the index using Callias' index theorem \cite{Callias}.
 The index of $\slashed{D}$ is given by
\begin{eqnarray}
\textrm{ind } \slashed{D} &=&\textrm{ind } \mathcal{L}_d =\lim_{z \to 0} \text{Tr}\,\left[\frac{z}{z+\mathcal{L}_d^\dagger \mathcal{L}_d} -\frac{z}{z+ \mathcal{L}_d\mathcal{L}_d^\dagger} \right]\nonumber\\
&=& -\frac{1}{2} \lim_{z \to 0}\sum_i \left[\frac{\xi_i(\tau \to \infty)}{(z+(\xi_i(\tau \to \infty))^2)^{1/2}} - \frac{\xi_i(\tau \to -\infty)}{(z+(\xi_i(\tau \to -\infty))^2)^{1/2}} \right]\nonumber \\
&=& -\frac{1}{2}\sum_i \left[\frac{\xi_i(\tau \to \infty)}{|\xi_i(\tau \to \infty)|} - \frac{\xi_i(\tau \to -\infty)}{|\xi_i(\tau \to- \infty)|} \right].
\label{indextheory1}
\end{eqnarray}
The $\tau$-dependence of $\xi_i$ is entirely contained in the function $\phi_{s/a}$, which takes values $0$ or $1$ at $\pm\infty$. Then
\begin{eqnarray}
\textrm{ind } \slashed{D}_{s/a} &=& \sum_{r=1}^{r_0}\left[-\frac{1}{2}(N_r-1)\left( \frac{\frac{3}{2} - \phi_{s/a}(\infty) \frac{N_r+1}{2}}{|\frac{3}{2} - \phi_{s/a}(\infty) \frac{N_r+1} {2}|}-\frac{\frac{3}{2} - \phi_{s/a}(-\infty) \frac{N_r+1}{2}}{|\frac{3}{2} - \phi_{s/a}(-\infty)\frac{N_r+1} {2}|}\right)\right.\nonumber\\
&&\left. -\frac{1}{2}(N_r+1)\left( \frac{\frac{3}{2} + \phi_{s/a}(\infty) \frac{N_r-1}{2}}{|\frac{3}{2} + \phi_{s/a}(\infty) \frac{N_r-1} {2}|}-\frac{\frac{3}{2} + \phi_{s/a}(-\infty) \frac{N_r-1}{2}}{|\frac{3}{2} + \phi_{s/a}(-\infty) \frac{N_r-1} {2}|}\right)\right].
\label{indextheory2}
\end{eqnarray}

As $0 \leq \phi_{s/a} (\tau) \leq 1$, the term in the second bracket of (\ref{indextheory2}) always evaluates to 
$0$, and the eigenvalues with degeneracy $(N_r+1)$ do not contribute to the index at all. 
The $\phi_{s/a}$ at $\tau = \pm \infty$ is given in (\ref{fsb}) and (\ref{fab}). Therefore, we have 
\begin{equation}
\text{ind } \slashed{D}_s = \frac{1}{2} \sum_{r=1}^{r_0} (N_r-1) \left[1- \frac{2-N_r}{|2-N_r|}\right]=-\text{ind } \slashed{D}_a,
\label{indextheory3}
\end{equation} 
where the subscripts $s$ and $a$ represent self-dual and anti-self-dual instanton backgrounds respectively.

For $N_r > 2$, the term in the square brackets in (\ref{indextheory3}) evaluates to $2$. Corresponding to each block of size $N_r \geq 3$, there exist $N_r-1$ normalizable zero modes of $\slashed{D}$ with one chirality ($+1$ for the self-dual instanton background and $-1$ for anti-self-dual one) and none with the other. This can be verified through examination of (\ref{sol2}) and (\ref{sol4}) at the boundaries.

When $N_r=1$, the index evaluates to $0$. Consequently there are no normalizable zero modes corresponding to the blocks of size $N_r=1$, i.e., the trivial representation.

\begin{figure}[!h]
\centering
\includegraphics[scale=1.1]{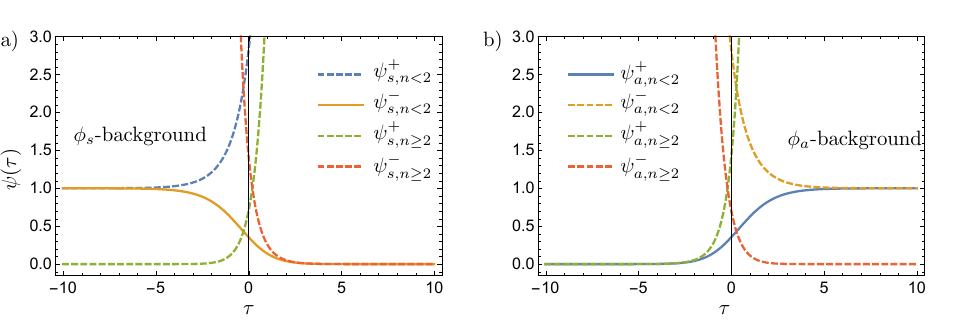}
\caption{Zero modes of  $\slashed{D}$ for  $N_r=2$ in  a)  a self-dual instanton, and  b) an anti-self-dual instanton background. Here, the solid lines are the resonance states, while the dashed lines are the non-normalizable states.}
\label{fig3}
\end{figure}
For the blocks with $N_r=2$, it seems at first sight that the index (\ref{indextheory3}) is undefined.
This is the case for which one eigenvalue of $\Sigma$ vanishes at either $\tau = -\infty$ or $+\infty$, depending on whether background is self-dual or anti-self-dual. Thus for large $|\tau|$, the Dirac operator resembles that of a free particle. The corresponding zero mode is not strictly normalizable (it is delta-function normalizable); it is a zero energy resonance, or a threshold state. The resonance states are plotted in 
Fig.\ref{fig3}c and Fig.\ref{fig3}d.

To summarize, if $L_i$'s consist of $r_0$ irreducible blocks with $r_1$ one-dimensional and $r_2$ 2-dimensional blocks, there are $r_2$ zero-energy resonances of one chirality and for each of the remaining $r_0 -r_1 -r_2$ blocks with dimension $N_r\geq 3$, there are $N_r-1$ normalizable zero modes of one chirality. The chirality is $+ 1$ for the self-dual background and $-1$ for the anti-self-dual background. As we shall see in the next section, both zero modes and zero energy resonance states have non-zero contribution to the axial anomaly. Therefore to correctly take into account all contributions, we must extend the definition of $\text{ind}\,\slashed{D}$ to include zero modes as well as zero energy resonances with positive and negative chirality, as in \cite{Bolle}. Then we can write down the index as
\begin{eqnarray}
\text{ind } \slashed{D}_s =-\text{ind } \slashed{D}_a = \sum_{r=1}^{r_0} (N_r-1) = N - r_0.
\label{index}
\end{eqnarray}
For the more general (anti-)self-dual instanton (\ref{ansatz}), the index is
\begin{eqnarray}
\text{ind } \slashed{D}_s =-\text{ind } \slashed{D}_a = r_0^{(2)} - r_0^{(1)}.
\label{index2}
\end{eqnarray}

This is the same as the new topological charge of the instanton background, obtained in (\ref{topcharge}). This is the matrix model version of the Atiyah-Singer index theorem:
\begin{equation}
\text{ind }\slashed{D} =\mathcal{T}_{\text{new}}.
\end{equation}
We emphasize that $\mathcal{T}_{\text{new}}$ is a quantity computed from the pure gauge sector, while the index  of the Dirac operator   counts the difference between the number of zero modes of opposite chiralities. It is remarkable that there exists such a simple relation between the two: a priori, there is no reason to expect this equality. Moreover, as we will show in Section 6, this relation between the charge and the index can be suitably adapted  for adjoint fermions as well, hinting towards its universal nature in the context of the matrix model.  However, despite the demonstration of the equivalence, there is no simple explanation of its origin and as it stands, the relation  $\mathcal{T}_{\text{new}}=$ ind $\slashed{D}$ is fortuitous.

\section{Non-invariance of fermion measure}

We adapt Fujikawa's method to demonstrate the axial anomaly, and relate it with the index of the Dirac operator. Specifically we show that under a $U(1)_A$ transformation, the measure of the fermionic path integral is not invariant. The Jacobian of the transformation gives the integrated anomaly.

The Euclidean fermionic path integral is given by
\begin{equation}
\int \mathcal{D}\bar{\Psi}\mathcal{D}\Psi e^{-S_E^F}.
\end{equation}

We expand the fermionic field $\Psi$ in the basis of the eigenfunctions of $\slashed{D}$ in a given background field configuration as
\begin{eqnarray}
\Psi = \sum a_n \Phi_n, \quad\quad \bar{\Psi} = \sum_n b_n \Phi_n, 
\end{eqnarray}
where $\Phi_n$'s include eigenfunctions of $\slashed{D}$ with non-zero eigenvalues, zero modes as well as zero-energy resonances. In this basis, the fermionic path integral measure is given by
\begin{equation}
d\mu\equiv\mathcal{D}\bar{\Psi}\mathcal{D}\Psi = \prod_{n,m} da_n db_m.
\end{equation} 
Under a $U(1)_A$ rotation with an infinitesimal $\alpha$, the transformed $\Psi'$ can be expanded as
\begin{eqnarray}
\Psi' = \sum a_n' \Phi_n, \quad\quad \bar{\Psi}' = \sum_n b_n' \Phi_n
\end{eqnarray}
where the coefficients transform linearly: 
\begin{eqnarray}
a_n' = C_{mn} a_m, \quad\quad b_n' = C_{mn} b_m. 
\end{eqnarray}
Therefore,
\begin{eqnarray}
\prod_n da_n' = [\det C_{nm}]^{-1} \prod_m da_m, \quad\quad \prod_n db'_n = [\det C_{nm}]^{-1} \prod_m db_m,
\end{eqnarray}
where
\begin{eqnarray}
[\det C_{nm}]^{-1} = e^{-i \int d\tau \, \alpha \mathcal{A}(\tau)}.
\end{eqnarray}
The anomaly function $\mathcal{A}(\tau)$ is defined as
\begin{eqnarray}
\mathcal{A}(\tau) = \sum_n \Phi_n^\dagger \gamma^5 \Phi_n.
\label{anomalyfn}
\end{eqnarray}
As a result, the path integral measure transforms as
\begin{eqnarray}
d\mu\rightarrow d\mu'&=&[\det C_{n'n}]^{-1} [\det C_{m'm}]^{-1} \prod_{n'} da_{n'} \prod_{m'} db_{m'}\nonumber\\
&=& e^{-2i \int d\tau \, \alpha \mathcal{A}(\tau)} d\mu. 
\label{meas_change}
\end{eqnarray}
Evaluating $\mathcal{A}(\tau)$ allows us to compute the change in the path integral measure and hence the anomaly. Now (\ref{anomalyfn}) is a summation over an infinite number of modes, and is formally divergent. We introduce a regulator $e^{- \beta \slashed{D}^2}$ and formally take the limit $\beta \to 0$ at the end:
\begin{eqnarray}
\mathcal{A}(\tau) = \lim_{\beta \to 0} \sum_n \Phi_n^\dagger \gamma^5 e^{- \beta \slashed{D}^2} \Phi_n.
\label{gamma5tr}
\end{eqnarray}

We already found that for non-zero eigenvalues, $\Phi_n$ and $\gamma^5\Phi_n$ are orthogonal, and so  eigenfunctions of $\slashed{D}$ with non-zero eigenvalues do not contribute to (\ref{gamma5tr}). However, the zero modes can have non-zero contribution to (\ref{gamma5tr}). Furthermore, the zero-energy scattering states, although not normalizable, can be delta-function normalized. Hence, they also have a well-defined contribution to (\ref{gamma5tr}).

Setting $\alpha=$ constant, the integrated anomaly from (\ref{meas_change}) is 
\begin{eqnarray}
 \int d \tau \, \mathcal{A} (\tau) &=& \int d \tau \lim_{\beta \to 0} \sum_n \Phi^\dagger_n \gamma^5 e^{-\beta \slashed{D}^2}\Phi_n\nonumber \\
&=&  \lim_{\beta \to 0} \int d \tau\,\left[ \sum_{k=1}^{n_+} \chi_k^{+ \dagger} \gamma^5 e^{-\beta \slashed{D}^2} \chi_k^+ +\sum_{k=1}^{n_-} \chi_k^{- \dagger} \gamma^5 e^{-\beta \slashed{D}^2} \chi_k^-
\right],
\end{eqnarray}
where $\chi_k^{\pm}$ include the zero modes as well as the zero-energy resonances. As $\slashed{D} \chi_k^\pm=0$, the summations are independent of $\beta$, and we can evaluate them at $\beta = 0$.
Therefore
\begin{eqnarray}
 \int d \tau \, \mathcal{A} (\tau) &=& \int d\tau \, \left[ \sum_{k=1}^{n_+} \chi_k^{+ \dagger} \gamma^5 \chi_k^+ + \sum_{k=1}^{n_-} \chi_k^{- \dagger} \gamma^5 \chi_k^- \right]. 
\end{eqnarray}
As $\chi_k^{\pm}$ are also eigenvectors of $\gamma^5$, the above implies  
\begin{eqnarray}
 \int d \tau \, \mathcal{A} (\tau)
&=& (n_+ -n_-)=\text{ ind } \slashed{D}.
\label{anomaly}
\end{eqnarray}

Thus in a background gauge configuration where the Dirac operator has a nonzero index, the fermion measure is not invariant under axial transformations, and axial symmetry is anomalously broken. Specifically, in the $SU(N)$ instanton background (\ref{ansatz}), $\text{ind }\slashed{D}$ is given by (\ref{index2}), and the integrated anomaly is given by
\begin{equation}
\alpha\int d \tau \, \mathcal{A}(\tau)= \pm\alpha (r_0^{(2)}-r_0^{(1)}). 
\end{equation}

Because $(r_0^{(2)}-r_0^{(1)})$ is always an integer,  $e^{\mp 2i \alpha \int \, d\tau\, \, \mathcal{A}(\tau)} = e^{\mp 2i\alpha (r_0^{(2)}-r_0^{(1)})} =1$ when $\alpha=n \pi$ for any $N$ (here 
$n \in \mathbb{Z}$). This means that under axial rotations with $\alpha= n \pi$, the anomaly vanishes. Thus, the $U(1)_A$ is anomalously broken to a discrete $\mathbb{Z}_2$ residual symmetry group.

The generalization to $N_f$ flavours of fermions is straightforward. 
The fermionic action is given by
\begin{equation}
S_E^F= \int d\tau \sum_{f,g=1}^{N_f}\bar{\Psi}_f \delta_{fg}(i\slashed{D})\Psi_g.
\end{equation}
The Dirac operator is diagonal in flavour, and we simply obtain $N_f$ copies of the spectrum of $\slashed{D} $ for a single flavour. The axial symmetry is now broken to $\mathbb{Z}_{2N_f}$.

\section{Adjoint Weyl Fermion} 

We consider here the case of adjoint Weyl fermions, which are relevant for supersymmetric gauge matrix models \cite{ErrastiDiez:2020iyk}. We show that in this case too, there is a non-trivial index of Dirac operator in the instanton background, and hence, an anomaly.

The Euclidean fermionic action with an adjoint Weyl fermion can be written as  
\begin{eqnarray}
S_E^F
&=& \int dt  \, \lambda^\dagger \left( \partial_\tau -A_0 + \sigma^i \mathcal{F}_i + \frac{3}{2} \right) \lambda
\end{eqnarray}
where $\mathcal{F}_i = M_{ia} G_{a}$ and $ G_a = -if_{abc}$ are the $SU(N)$ generators in the adjoint representation: 
\begin{eqnarray}
\mathcal{F}_i = M_{ia} G_{a} = 2 \text{ Tr}(A_i T^a) G_a. 
\end{eqnarray}
As before, we choose the temporal gauge.

\subsection{Embedding the Instanton in $(N^2-1)$ Dimensions}

The matrix $\mathcal{F}_i$ is an $(N^2-1)$-dimensional matrix, and is just the embedding of the background gauge field $A_i$ in $N^2-1$ dimensions. In particular if $A_i = \phi_{s/a} L^{(1)}_i +(1-\phi_{s/a})L^{(2)}_i$, $\mathcal{F}_i$ takes the form
\begin{equation}
\mathcal{F}_i = \phi_{s/a} \mathcal{J}^{(1)}_i + (1-\phi_{s/a}) \mathcal{J}_i^{(2)}.
\end{equation}
where $\mathcal{J}_i^{(\alpha)}$ are the representations of $SU(2)$ obtained by embedding $L_i^{(\alpha)}$ in $(N^2-1)$ dimensions.

To understand this embedding, we recall that adjoint representation of colour is derived from the tensor product of the fundamental and the anti-fundamental representation:
\begin{equation}
N\otimes\bar{N}=(N^2-1)\oplus \mathbb{1}.
\label{group}
\end{equation} 
Since $L_i$ and $\mathcal{J}_i$ are embeddings of the $SU(2)$ Lie algebra into the $N$-dimensional and the $(N^2-1)$-dimensional representations of $SU(N)$ respectively, (\ref{group}) induces a map between the embeddings as well. If $L_i$ is the $N$-dimensional irreducible representation with spin-$j=\frac{N-1}{2}$, the induced map gives the angular-momentum algebra:
\begin{equation}
j\otimes j = 2j\otimes (2j-1) \otimes...\otimes 0.
\label{spinsum}
\end{equation}
On the other hand, if $L_i$ is a reducible representation made of $r$ blocks with spin $j_r=\frac{N_r-1}{2}$ in each block, we obtain
\begin{equation}
\left(\oplus_{r} j_r\right)\otimes \left(\oplus_q j_q\right) = \bigoplus_{r,q} \big[(j_r+ j_q)\oplus (j_r+ j_q-1)\oplus \cdots \oplus (|j_r-j_q|)\big]. 
\end{equation}
In the above direct sum, one spin-0 block arises from the singlet $\mathbb{1}$ of the RHS in (\ref{group}), elimination of which gives us $\mathcal{J}_i$. So for $L_i$ in a representation $\sum_r N_r$, $\mathcal{J}_i$ is given by one singlet representation removed from the following direct sum:
\begin{equation}
\bigoplus_{r,q} [(N_r+N_q-1)\oplus (N_r+N_q-3)\oplus ...\oplus  (|N_r-N_q|+1)] .
\label{reps}
\end{equation}
Thus $\mathcal{J}_i^{(\alpha)}$ is always a direct sum of $q_0^{(\alpha)}$ irreps of dimensions $\mathcal{N}_q^{(\alpha)}$, with $\sum_{q=1}^{q_0^{(\alpha)}} \mathcal{N}_q^{(\alpha)} = N^2-1$. We can again define the spin in each block as $j_q^{(\alpha)} = \frac{\mathcal{N}_q^{(\alpha)}-1}{2}$, and the electric and magnetic fields with respect to the embedding $\mathcal{F}_i$ as
{ \begin{eqnarray}
\mathcal{E}_i = \frac{\partial \mathcal{F}_i}{\partial \tau},\quad \quad \mathcal{B}_i = -\mathcal{F}_i-\frac{i}{2}\epsilon_{ijk}[\mathcal{F}_j,\mathcal{F}_k]
\end{eqnarray}}
which satisfy the instanton equation $\mathcal{E}_i=\pm \mathcal{B}_i$. Then the quantity
\begin{eqnarray}
\mathcal{T}^{\text{adj}} &=& 4\int d\tau \text{Tr }\mathcal{E}_i\mathcal{B}_i =\pm \frac{2}{3} 
\left(\text{Tr}\mathcal{J}_i^{(2)}\mathcal{J}_i^{(2)} - \text{Tr}\mathcal{J}_i^{(1)}\mathcal{J}_i^{(1)} \right) \\
&=& \pm\frac{2}{3} \left(\sum_{q=1}^{q_0^{(2)}} j_q^{(2)}(j_q^{(2)}+1)(2j_q^{(2)}+1) - \sum_{q=1}^{q_0^{(1)}} j_q^{(1)}(j_q^{(1)}+1)(2j_q^{(1)}+1) \right)
\end{eqnarray}
has the same form as (\ref{instantoncharge}). 

To express $\mathcal{T}^{\text{adj}}$ in a more convenient form like before, we define $\mathcal{E'}_i$ exactly as in (\ref{Erescaled}) as
\begin{eqnarray}
\mathcal{E'}_i = \frac{d \phi_{s/a}}{d \tau}(e_i^{(2)} - e_i^{(1)}), \quad e_i^{(\alpha)} = \bigoplus_{q=1}^{q_0^{(\alpha)}}\frac{3}{(j_q^{(\alpha)}+1)(2j_q^{(\alpha)}+1)}\mathcal{J}_i^{(\alpha),q}
\end{eqnarray}
where $\mathcal{J}_i^{(\alpha),q}$ are the irreducible blocks with spin $j_q^{(\alpha)}$ in $\mathcal{J}_i^{(\alpha)}$.
Then, the new instanton charge for the embedding is given by
 \begin{eqnarray}
\mathcal{T}^{\text{adj}}_{\text{new}}&=&\int d\tau \text{Tr }\mathcal{E'}_i\mathcal{B}_i = q_0^{(2)}-q_0^{(1)}.
\end{eqnarray}

\subsection{Index of the Dirac Operator}

The index calculation is exactly as in Section 4 from (\ref{indextheory1}) and (\ref{indextheory2}), by replacing $N_r^{(\alpha)}$ by $\mathcal{N}_q^{(\alpha)}$ with $\sum_q \mathcal{N}_q^{(\alpha)}=N^2-1$. By correctly taking into account the zero modes as well as the zero-energy resonances, the index is now given by
\begin{equation}
\text{ind }\slashed{D}  = q_0^{(2)}-q_0^{(1)} =\mathcal{T}^{\text{adj}}_{\text{new}}.  \label{index_adjoint}
\end{equation}

Furthermore, we observe that all the even-dimensional blocks occurs an even number of times. This is understandable because such a representation can come only from the cross terms in (\ref{reps}) (for which $r\neq q$), which occur twice in the direct sum. So all distinct eigenvalues occur an even number of times, i.e., are at least doubly degenerate. This can also be seen as follows. Given
\begin{equation}
\left(\sigma_i\otimes\mathcal{F}_i +\frac{3}{2}\mathbf{1} \right)\varphi = \lambda \varphi,
\end{equation}
taking the complex conjugate and simplifying using $\sigma_i^*=-\sigma_2\sigma_i\sigma_2$ and $\mathcal{F}_i^*=\mathcal{F}_i$ gives
\begin{eqnarray}
\left(\sigma_i\otimes\mathcal{F}_i +\frac{3}{2}\mathbf{1} \right)(\sigma_2\otimes\mathbf{1})\varphi^* = \lambda(\sigma_2\otimes\mathbf{1}) \varphi^*. 
\end{eqnarray}
Thus corresponding to an eigenvector $\varphi$, we have a degenerate eigenvector $(\sigma_2\otimes\mathbf{1})\varphi^*$. The index is always an even integer, and so the anomaly $e^{-2i\alpha\text{ ind}\slashed{D}}$ takes value 1 when $\alpha=\frac{n\pi}{2}$, $n\in \mathbb{Z}$. Therefore a single adjoint Weyl fermion breaks the $U(1)_A$ axial symmetry to a residual $\mathbb{Z}_4$  subgroup.  

For $N_f$ flavours, the residual symmetry is $\mathbb{Z}_{4N_f}$.

\section{Discussion}

As we mentioned in the introduction, the gauge matrix model is very different from the corresponding gauge field theory. Nevertheless, the matrix model does retain important non-perturbative features of the full field theory. As we have 
demonstrated here, the axial anomaly is one such feature. In the usual discussion of the axial anomaly in non-Abelian gauge field theories, only the irreducible connections are considered, and it is the instanton number of such connections that is related to the fermion zero modes. A priori, there is no reason that the residual axial symmetry should match with the corresponding result in the field theory, and it is surprising that it matches for 
the case of fundamental fermions. Whether this coincidence has a deeper reason requires further investigation.

Our result on the anomaly provides a strong conceptual support to the numerical investigations of the matrix model \cite{Pandey:2019dbp}. In addition to reproducing the masses of light hadrons with excellent accuracy, the numerics also 
show that the pseudo-scalar mesons are much lighter than their scalar counterparts. Furthermore, it also finds the $\eta'$-meson to be considerably heavier than the $\eta$-meson. The result on the axial anomaly presented in this article serves to 
strengthen the position of the matrix model as an effective low-energy approximation of QCD.

There is a plethora of matrix models that have been studied in the literature \cite{Ishibashi:1996xs,Banks:1996vh,Berenstein:2002jq} that have a non-Abelian gauge symmetry and fermions, and which remain a subject of continuing interest \cite{Hadizadeh:2004bf, Costa:2014wya, Filev:2015hia, Asano:2018nol, Asplund:2015yda, Han:2019wue}. Whether axial symmetry continues to hold at the quantum level in these models needs to be understood, in light of the results presented here. 
 
Lastly, the axial anomaly is present for any $SU(N)$ gauge group, and there is no reason to expect that it is washed out in the large $N$ 
limit.

\noindent {\bf Acknowledgments:} We are grateful to Denjoe O'Connor and V. Parameswaran Nair for suggesting the use of Callias' Index Theorem and useful discussions.


\end{document}